\documentclass[twocolumn,nolinenumbers]{aastex631}

\newcommand{\logg} {\log \textsl{\textrm{g}}}
\usepackage{xcolor,tikz}
\usepackage{gensymb}
\usepackage{multirow}
\usepackage{lineno}
\usepackage{float}

\shorttitle{Precise Parameters for Two LISA Sources}
\shortauthors{Barrientos et al.}

\begin{document}

\title{Precise Parameters for Two LISA Sources}

\thanks{Author e-mail: \href{mbarrientos@ou.edu}{mbarrientos@ou.edu}}

\author[0000-0002-6153-9304]{Manuel Barrientos} 
\affiliation{Homer L. Dodge Department of Physics and Astronomy, University of Oklahoma, 440 W. Brooks St., Norman OK, 73019, USA}

\author[0000-0001-6098-2235]{Mukremin Kilic} 
\affiliation{Homer L. Dodge Department of Physics and Astronomy, University of Oklahoma, 440 W. Brooks St., Norman OK, 73019, USA}

\author[0000-0002-4462-2341]{Warren R.\ Brown}
\affiliation{Center for Astrophysics, Harvard \& Smithsonian, 60 Garden Street, Cambridge, MA 02138, USA}

\author{Fatma Ben Daya}
\affiliation{Hamburger Sternwarte, Univeristy of Hamburg, Gojenbergsweg 112, 21029 Hamburg, Germany}

\author[0000-0002-2384-1326]{Antoine B{\'e}dard}
\affiliation{Department of Physics, University of Warwick, CV4 7AL, Coventry, UK}

\author[0000-0002-9574-578X]{Tyson Littenberg}
\affiliation{NASA Marshall Space Flight Center, Huntsville, Alabama 35812, USA}

\author[0000-0002-6540-1484]{Thomas Kupfer}
\affiliation{Hamburger Sternwarte, Univeristy of Hamburg, Gojenbergsweg 112, 21029 Hamburg, Germany}

\author[0000-0002-0801-8745]{Snehalata Sahu}
\affiliation{Department of Physics, University of Warwick, CV4 7AL, Coventry, UK}

\begin{abstract}
We present precise parameters for two compact double white dwarf binaries, SDSS J232230.20+050942.0 (J2322+0509) and SDSS J063449.92+380352.2 (J0634+3803), 
with orbital periods of 20 and 26.5 minutes, respectively. These systems will serve as verification sources for the Laser Interferometer Space Antenna (LISA).
To significantly improve the electromagnetic (EM) constraints on these two systems and the LISA detectability predictions, we conducted spectroscopic follow-up observations using HST/STIS, Keck I/LRIS, and Keck II/ESI. Our analysis significantly improves the temperature, surface gravity, and mass constraints for both primary and secondary components in J2322+0509, as well as dynamical properties such as radial velocities and orbital periods in both systems. For J2322+0509, we derive an updated inclination of $i$ = 25$^{+4.5}_{-3.0}$  deg, while for J0634+3803, we obtain $i$ = 43$^{+7.0}_{-5.6}$  deg. We assess the detectability of these sources using LDASOFT. Incorporating EM priors on inclination significantly enhances the gravitational wave signal recovery, reducing uncertainties in amplitude by a factor of 2-4 and shortening the detection time by up to a few months. Our results underscore the importance of multi-messenger observations in characterizing double white dwarf binaries and maximizing LISA's early scientific capabilities.
\end{abstract}

 
\section{Introduction} 
\label{sec:1}

Double white dwarf (DWD) systems are predicted to be the most common/numerous binaries in our Galaxy \citep[e.g.,][]{nelemans2001,nissanke2012,korol2017}. Some of the fastest DWD binaries have orbital periods as short as several minutes \citep[e.g.,][]{brown2011,burdge2020,as2023}. These systems are prime candidates for detection by the upcoming space-based gravitational wave (GW) detectors in the mHz frequency range, e.g., the Laser Interferometer Space Antenna \citep[LISA;][]{as2017}, which will open a new window into the study of ultracompact binaries. 

Given their ``multi-messenger'' nature, these DWD binaries can also be observed electromagnetically, which offers a unique chance to characterize the Galactic DWD population by measuring their system parameters using both methods \citep[e.g.,][]{korol2019,li2020}. For instance, electromagnetic (EM) data can help constrain the position and binary inclination, which can help reduce the uncertainties in the GW amplitude measurements \citep{shah2012,shah2013}. Thanks to Gaia astrometry \citep{gaiadr3}, more precise distance measurements are also available; those can considerably improve the constraints on the chirp mass  \citep{s&n2014} and help their detectability with LISA \citep{kupfer2024}.

Multi-messenger detached DWD systems were first discovered by the Extremely Low-Mass (ELM) Survey \citep[][and references therein]{brown2010,brown2022,kosa2023} through time-series radial velocity measurements of ELM WD candidates. The ELM Survey discovered six DWD binaries with orbital periods $\lesssim$30 minutes \citep[e.g.,][]{brown2011,brown2020,kilic2014,kilic2021,kosa2025}. Recently, \citet{burdge2019,burdge2020,burdge2023} and \citet{chickles2025} have discovered 11 short-period binaries with $P\lesssim30$ min through a photometric search of the Zwicky Transient Facility \citep[ZTF;][]{ztf2019} and the Transiting Exoplanet Survey Satellite  \citep[TESS;][]{tess2015} data, increasing the sample of EM+GW DWD binaries to more than 50 sources \citep[][]{kupfer2024}.

However, some of these sources do not have precise EM constraints, which may impact their detectability in GW, especially for sources close to the LISA Signal-to-Noise (SNR) threshold \citep[][see their Section~3]{finch2023}. In particular, SDSS J232230.20+050942.0 \citep[hereafter J2322+0509,][]{brown2020} and SDSS J063449.92+380352.2 \citep[hereafter J0634+3803,][]{kilic2021} are two systems where EM parameters can be significantly improved with new observations, especially considering that both binaries have nearly face-on inclinations.

J2322+0509, located at a distance of 865 pc, is a single-lined spectroscopic binary with a 20-minute orbital period that contains two He-core WDs. \citet{brown2020} reported T$_{\rm{eff}}$= 19000 $\pm$ 1000 K and $\logg$= 7.0 $\pm$ 0.15 for the primary, and T$_{\rm{eff}}$= 8000 $\pm$ 1500 K and $\logg$= 7.0 $\pm$ 0.25 for the secondary, with masses of 0.27 and 0.24 M$_{\odot}$, respectively. They obtained a binary inclination of around $27\degree$, indicating a nearly face-on and non-eclipsing system. The optical data on this system shows about 15\% flux contribution from the cooler secondary WD, limiting the precision of the model atmosphere analysis on the primary WD.

J0634+3803 is a closer detached DWD binary with $g= 17$ mag at a distance of $\approx$435 pc \citep[][]{bj2021}. \citet{kilic2021} reported T$_{\rm{eff}}$= 27300 K and $\logg$= 7.46 dex for the primary, and T$_{\rm{eff}}$= 10500 K and $\logg$= 6.72 dex for the secondary, with masses of 0.45 and 0.21 M$_{\odot}$, respectively. However, because of the flux contribution from the cooler secondary, they found degeneracies in the resulting stellar parameters for the primary and secondary WDs (see their Figure 6). Furthermore, their reported orbital parameters for this system had relatively large uncertainties, e.g., $P=1591.4 \pm 28.9$ s. 

As part of our efforts to better characterize short-period DWD binaries in the mHz frequency range, we performed spectroscopic follow-up observations of J2322+0509 and J0634+3803 with HST/STIS, Keck I/LRIS, and Keck II /ESI. Here, we use these observations to improve the orbital and physical parameters of both components in each system. We present our observations and analysis for J2322+0509 and J0634+3803 in Section~\ref{sec:2} and \ref{sec:3}, respectively. We discuss the implications of using our new EM priors in LISA simulations, their GW characteristics, and their detectability in the first few months of observations in Section~\ref{sec:4}. We conclude in Section~\ref{sec:5} by highlighting the main findings and future work.

\begin{figure*}
\centering
\includegraphics[width=6in]{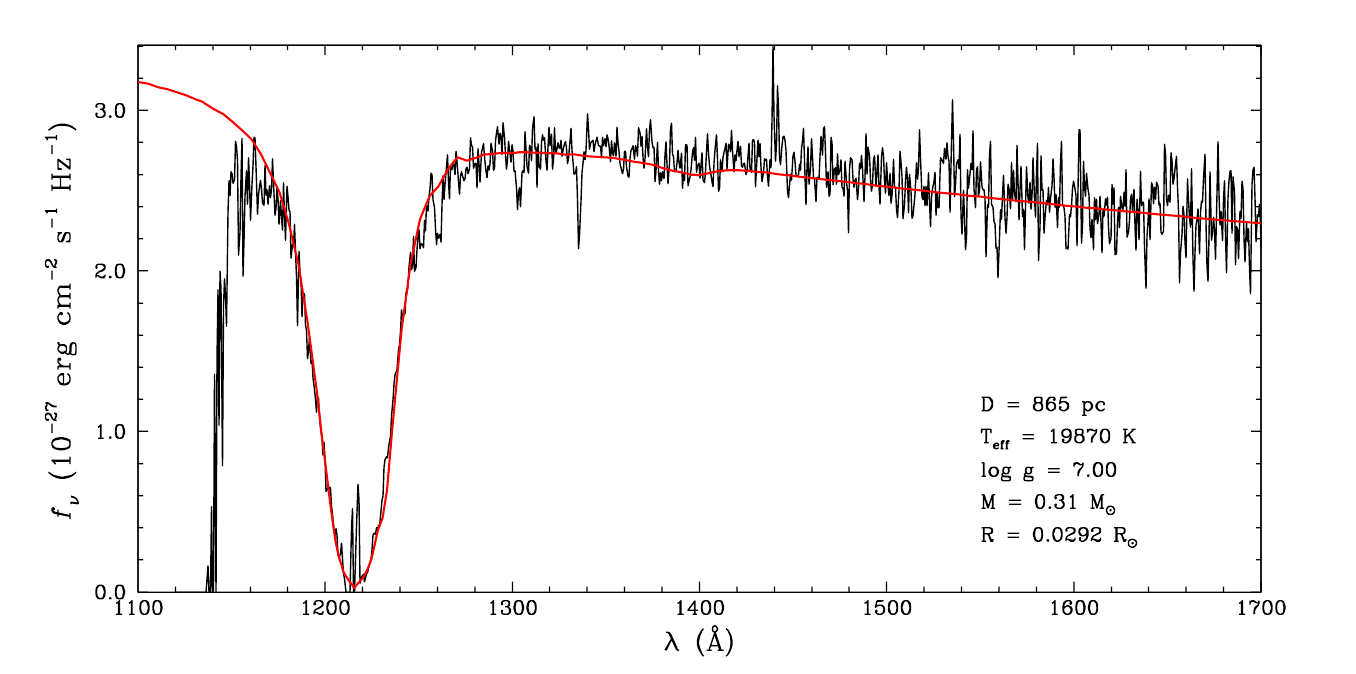}
\includegraphics[width=6in]{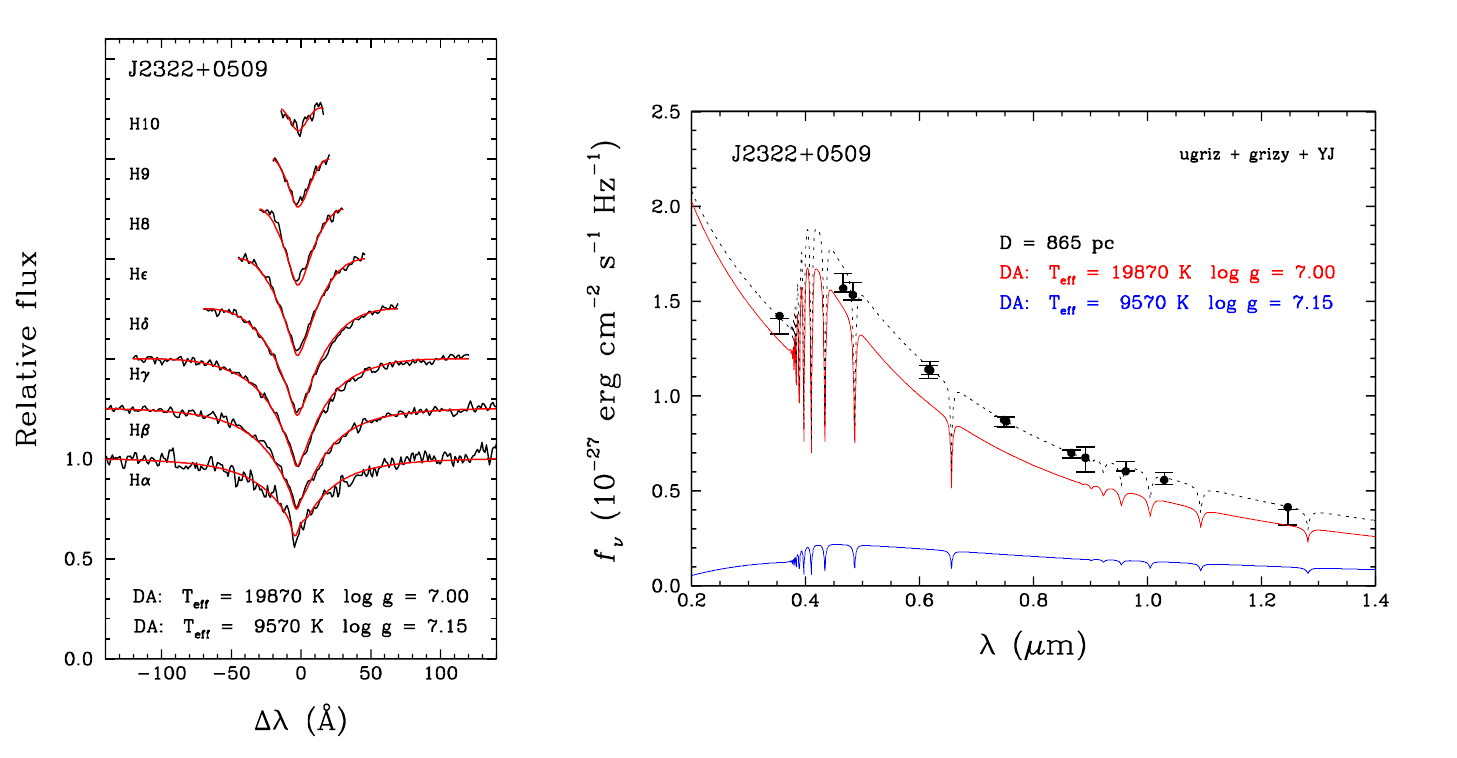}
\caption{Top panel: The HST/STIS spectrum of J2322+0509 (black line) along with the best-fitting DA model (red line). The narrow lines redward of Ly$\alpha$ are interstellar lines (Si\,\textsc{II} at 1260 \AA, O\,\textsc{I} at 1302 \AA, and C\,\textsc{II} at 1335 \AA). Bottom panels: Joint fit to the optical spectrum (left) and photometry (right) for the best DA + DA model. Here, the parameters of the primary are set to the results obtained from the HST/STIS spectrum. The left panel shows the synthetic model (red) overplotted on the observed spectrum (black). The right panel shows the synthetic fluxes (filled circles) and observed fluxes (error bars). The red and blue lines show the contribution of each WD to the total monochromatic model flux, displayed as the black dotted line.}
\label{fig:1}
\end{figure*}

\section{SDSS J232230.20 + 050942.0}
\label{sec:2}

\subsection{Physical Parameters}
\label{sec:2.1}

To significantly improve the constraints on the physical parameters of the 20-minute orbital period detached binary J2322+0509, we obtained UV spectra over 4 orbits with HST/STIS and the G140L grating as part of the program 16286 (PI: Kilic) in cycle 28. The G140L grating provides spectral coverage from 1150 to 1730 \text{\AA} with a resolving power of R$\sim$1000. One of the four orbits failed due to guide star re-acquisition problems. The exposure times for the three completed orbits are 2205, 2725, and 2725 s, respectively. The reduced spectra were retrieved from the Mikulski Archive for Space Telescopes (MAST) and combined using standard \texttt{IRAF} routines. These data can be accessed via the doi: \href{http://dx.doi.org/10.17909/35rz-vb77}{10.17909/35rz-vb77}.

Figure~\ref{fig:1}, top panel, shows our combined dereddened HST/STIS spectrum of J2322+0509 (black line) along with our best-fitting atmosphere model. The UV spectrum is dominated by the hot WD primary, with negligible contribution from the cooler secondary (see below). We use this spectrum to constrain the parameters of the primary WD. Our analysis uses the pure-H atmosphere models of \citet{tremblay2011} and the He-core mass-radius relation of \citet{althaus2013}. The spectrum is dereddened assuming $E(B-V) = 0.062$ from \citet{lallement2022} and the reddening law of \citet{gordon2023}. The STIS spectrum is best-fit by a model with $T_{\rm eff} = 19870 \pm 300$ K, $\log{g} = 7.00 \pm 0.10$, and $M = 0.31 \pm 0.02~M_{\odot}$. With these new UV constraints, parameter degeneracies no longer dominate the errors. However, the error in surface gravity is still relatively large due to the imprecise {\emph Gaia Data Release 3}  (DR3) parallax with $\sim$20\% uncertainty. 

\begin{figure}
\centering
\includegraphics[width=3.2in]{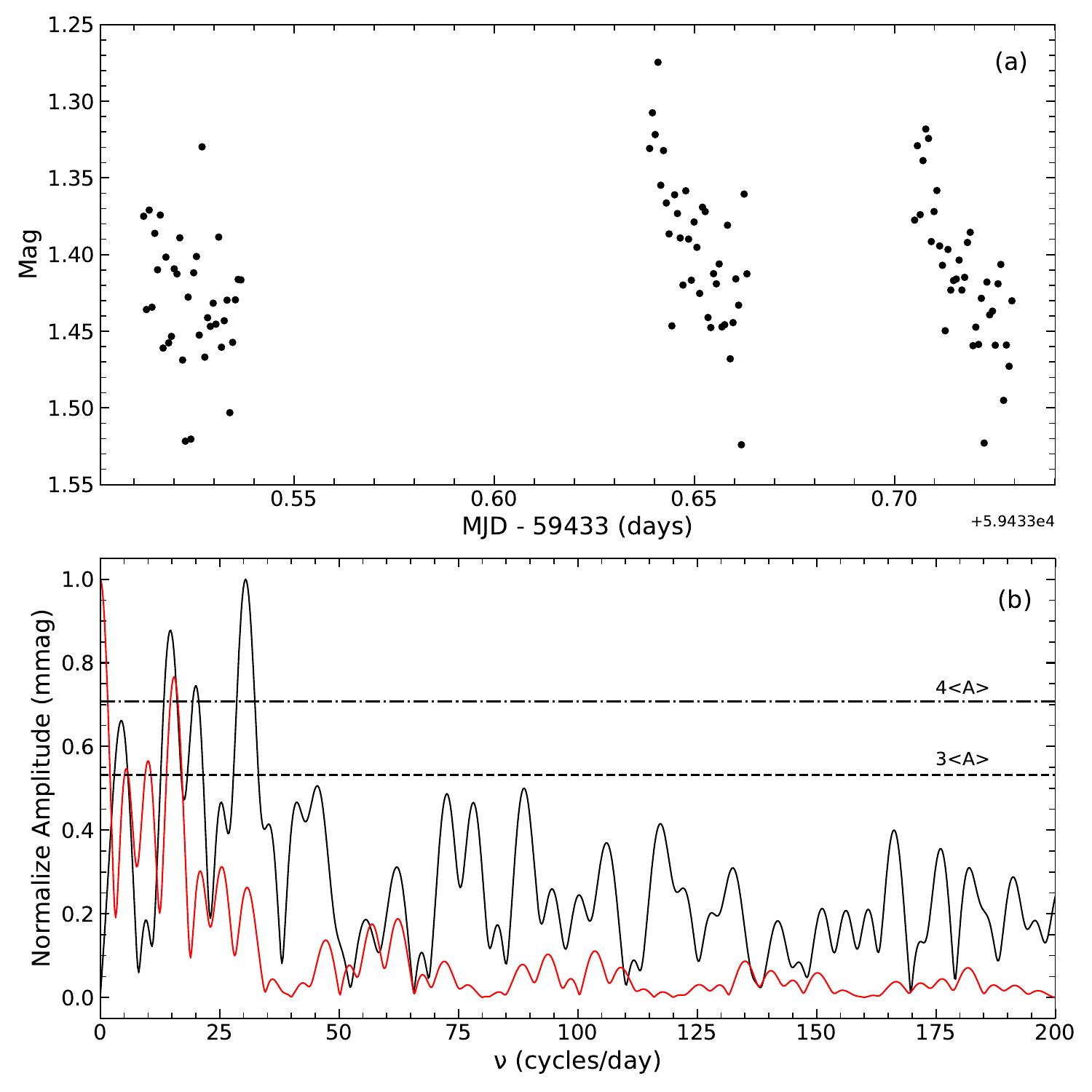}
\includegraphics[width=3.2in]{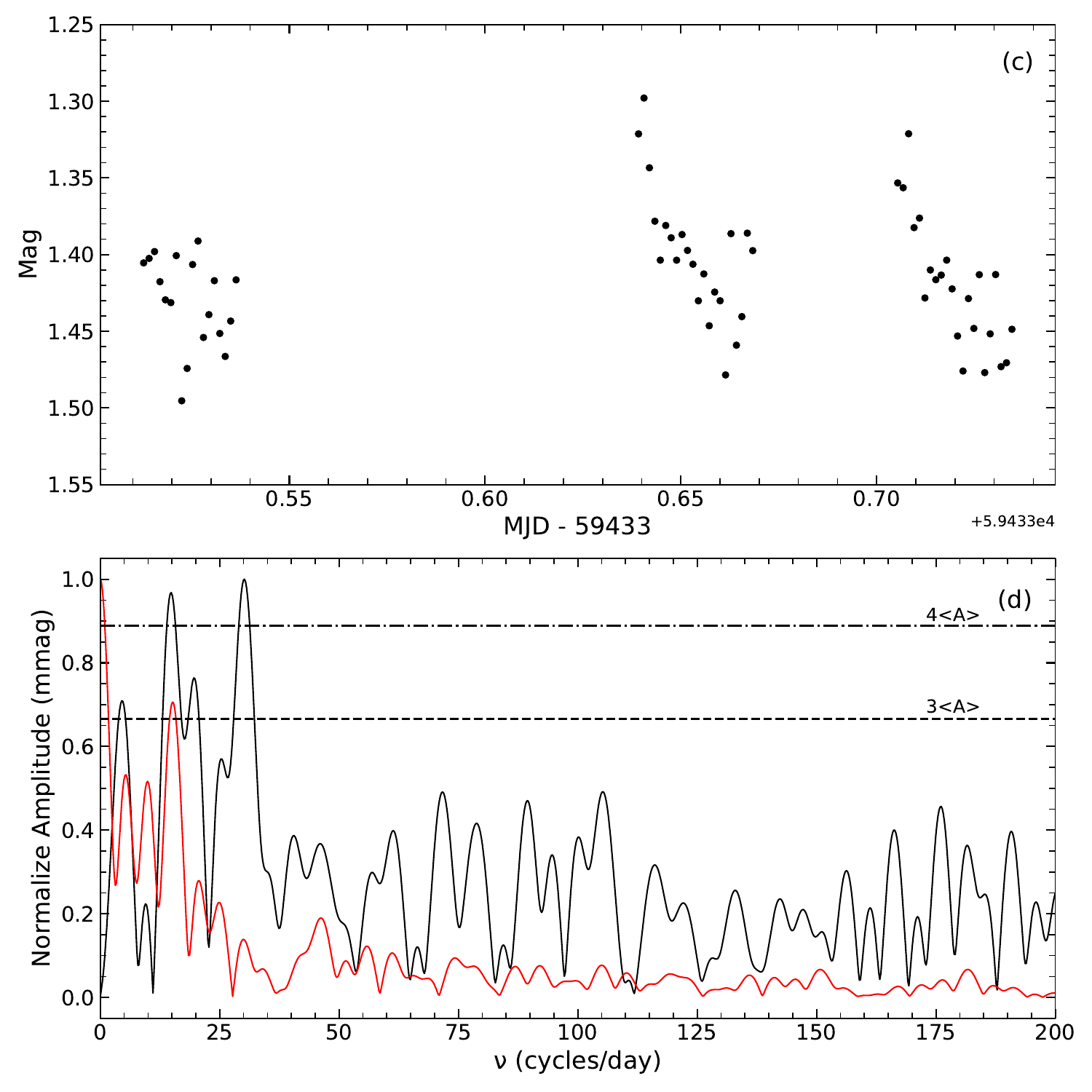}
\caption{HST/STIS light curves of J2322+0509 for 60 seconds and 120 seconds integration times, panel (a) and panel (c), respectively. Panels (b) and (d) show the Fourier transform of each light curve (black) along with the spectral window (red). The dotted and dashed-dotted lines mark the 3 and 4$<$A$>$ levels, where $<$A$>$ is the average amplitude in the Fourier transform.}
\label{fig:2}
\end{figure}

\begin{figure}
\centering
\includegraphics[width=3.2in, clip=true, trim=0.3in 2.2in 0.4in 1.4in]{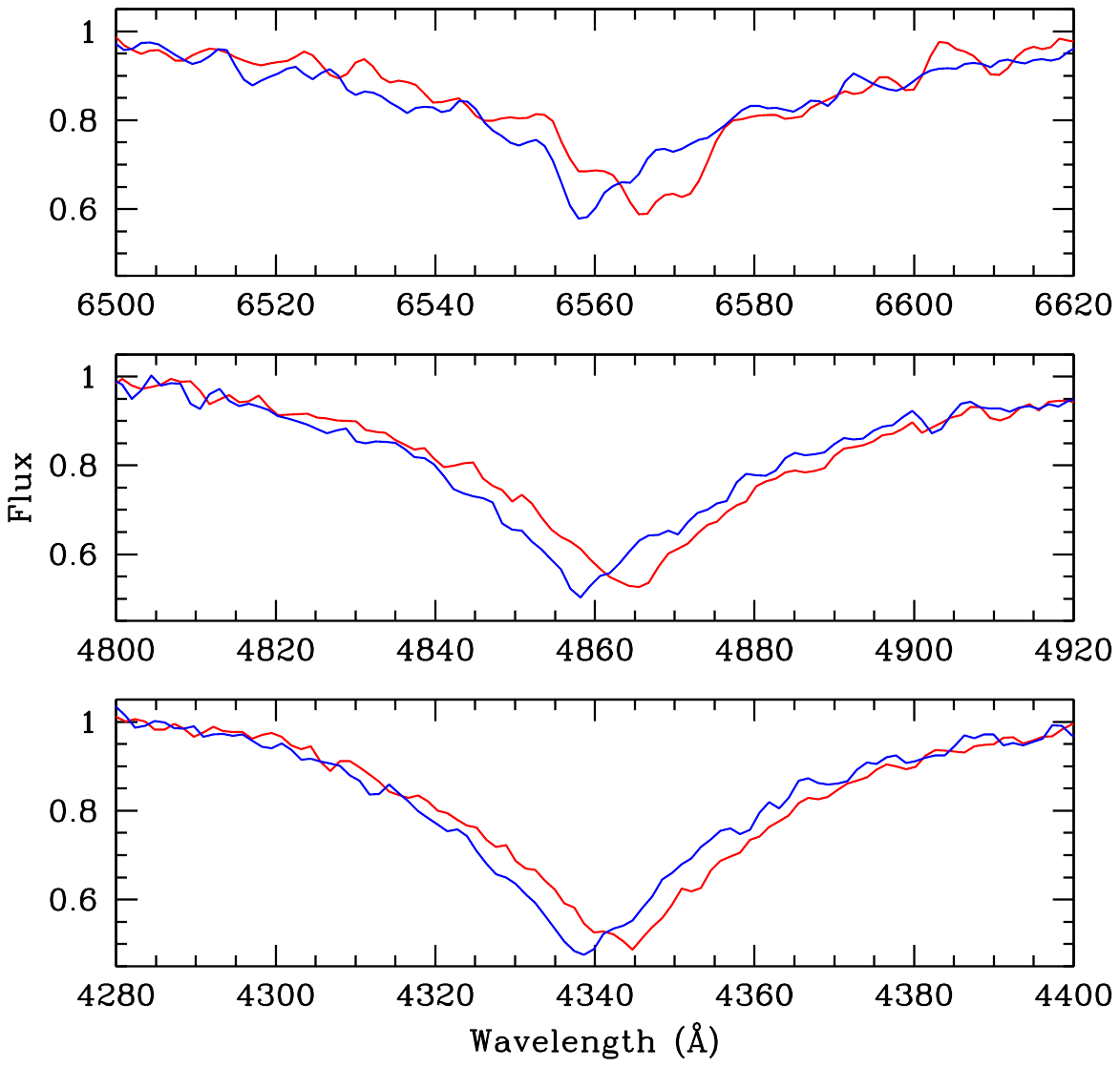}
\caption{Composite Keck I/LRIS spectra for J2322+0509  during negative (blue line) and positive (red line) quadratures. H$\alpha$, H$\beta$, and H$\gamma$ are shown in the top, middle, and bottom panels, respectively. H$\alpha$ spectra shown in the top panel are smoothed by three pixels for display purposes.}
\label{fig:3}
\end{figure}

We now perform a joint analysis that simultaneously considers our spectroscopic, astrometric, and photometric constraints. Our approach is to construct composite DA+DA binary WD models by adding two synthetic WD spectra, properly weighted by their respective radii. We simultaneously fit the spectroscopic Balmer line profiles and the broadband photometric
measurements using the astrometric parallax constraint (see \citealt{bedard2017} and \citealt{kilic2020} for details). We use dereddened SDSS ugriz \citep[][]{sdss2018}, Pan-STARRS grizy \citep[][]{panstarrs2012,panstarrs2016}, and UKIDSS YJ photometry \citep[][]{ukidss2007}. We fix the parameters of the hot WD based on our model fit to the HST/STIS spectrum. Fixing these parameters, we obtain $T_{\rm eff} = 9570 \pm 800$ K, $\log{g} = 7.15 \pm 0.20$, and $M = 0.29\pm0.05 ~M_{\odot}$ for the secondary star. Figure~\ref{fig:1}, bottom panels, show the results from this joint fit to the Balmer line profiles and the overall spectral energy distribution. This DA+DA WD model provides an excellent match to all available optical and UV data.

\subsection{Orbital Parameters}
\label{sec:2.2}

\citet{brown2020} presented time-series optical photometry of J2322+0509 and ruled out variability at the 0.35\% level. They reported tentative evidence of a peak in the Fourier Transform at the orbital frequency, but this peak was below their detection threshold. Even though there is no evidence of variability, they quantified the predicted Doppler beaming amplitude using the formalism from \citet{shporer2010} which yields a contribution of $\sim$0.1\% and ruled out any ellipsoidal variations. Both effects are strongly dependent on the system's physical parameters: Doppler beaming scales with the stellar temperature and radial velocity semi-amplitude, while ellipsoidal variations depend primarily on the mass ratio, primary radius, and orbital inclination \citep{M&N1993}.

To investigate potential UV variability, we analyzed our HST/STIS TIME-TAG mode observations, extracting light curves with 60 and 120 s integration times. Figure~\ref{fig:2} shows these light curves and their corresponding Fourier Transforms. Given the observing window over four orbits (with failed observations in the second orbit), the Fourier transforms do not reveal any significant variability at the binary's orbital period. Hence, the UV time-series photometry does not provide any new constraints on the orbital parameters of this binary. Nonetheless, UV and optical observations can provide complementary constraints to the system. The UV light curves are more sensitive to temperature-dependent variations and effectively isolates flux from the hotter component, thereby minimizing signal cancellation in Doppler beaming analysis. In contrast, the optical light samples both stars with potentially better phase coverage and SNR. Despite this, the lack of detectable modulation in either case supports a low-inclination, detached binary configuration.

We obtained additional follow-up optical spectroscopy of the same binary using Keck I/LRIS as part of the program N099 on UT 2020 Oct 20. We used a 1 arcsec slit and the 600 line mm$^{-1}$ grating in the blue and the 900 line mm$^{-1}$ grating in the red arm. We obtained 120-150 s long back-to-back exposures of this system over 4 hours to look for evidence of an H${\alpha}$ line from the cooler secondary star. 

Figure~\ref{fig:3} shows the composite spectra for J2322+0509 during negative (minimum radial velocity) and positive (maximum radial velocity) quadrature for the strongest H lines in the blue and red, respectively. We do not see any clear absorption features from the secondary star in any of the H lines. We follow the same approach as in \cite{brown2020} to obtain the radial velocity measurements. In brief, we cross-correlate the observed data with a summed, rest-frame template of the same target. Then, we do a $\chi^2$ minimization considering a circular orbit and account for phase smearing during our 120-150 s long exposures. The internal uncertainties are calculated by bootstrapping the measurements. 

\begin{figure}
\centering
\includegraphics[width=3.2in, clip=true, trim=0.3in 2.5in 0.6in 1.4in]{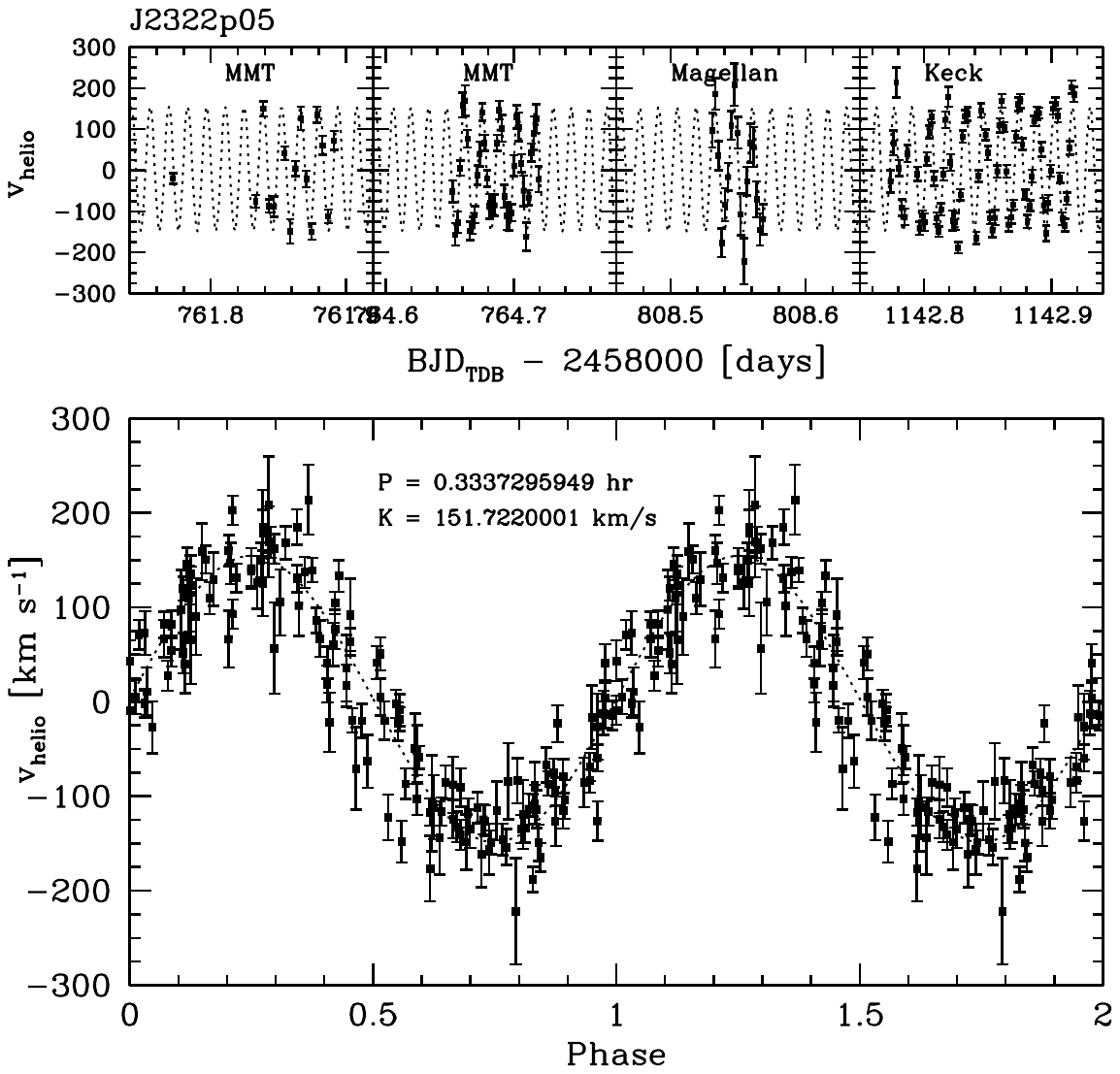}
\caption{Radial velocities of J2322+0509  based on all available data. The dotted line shows the best-fitting model for a circular orbit. The bottom panels show all data points phased with the best-fit period.}
\label{fig:4}
\end{figure}

Figure~\ref{fig:4} shows our radial velocity measurements along with the best-fitting circular orbit for all available data from \cite{brown2020} and including our new Keck I/LRIS measurements. Combining the entire set of data from the MMT, Magellan, and Keck, the best-fitting circular orbit has $P = 0.0139054 \pm 0.0000666$ d = $1201.4 \pm 5.8$ s, $K = 151.7 \pm 5.8$ km s$^{-1}$, and $\gamma = 3.1 \pm 4.3$ km s$^{-1}$. These results are consistent with \cite{brown2020} within the errors, with slight improvements in the uncertainties. 

We use the mass function and the improved physical and orbital parameters to precisely constrain the binary inclination as follows

\begin{equation}
    \sin(i) = K \left(\frac{P}{2\pi G}\right)^{1/3} \frac{(M_1 + M_2)^{2/3}}{M_2},
    \label{eq:1}
\end{equation}

\noindent where $i$ is the binary inclination, $P$ is the orbital period, $K$ is the velocity semi-amplitude, and M$_1$ and M$_2$ are the masses for the primary and secondary components, respectively. With the improved constraints, we obtain an inclination of $i= 25^{+4.5}_{-3.0}$ deg. Adopting the new parameters and applying Equation~4 from \citet{shporer2010}, we estimate a Doppler beaming amplitude of 0.096\% which is in agreement with \citet{brown2020}. Table~\ref{tab:1} summarizes the revised system parameters for J2322+0509. 

\section{SDSS J063449.92+380352.2}
\label{sec:3}


To better constrain the radial velocity and orbital period for J0634+3803, we obtained follow-up spectra with the Echellette Spectrograph and Imager \citep[ESI;][]{esi2002} on the 10m Keck II telescope on Mauna Kea on UT 2022 Feb 6 as part of the program N024. We used ESI echelle mode with a 0.75 arcsec slit and $2\times2$ binning to obtain spectra with a resolving power of 5400 (55.9 km s$^{-1}$ per resolution element) over the wavelength range from 3900-10000 \AA. With 120-second long exposures, we obtained a SNR of around 14 per pixel at H$\alpha$ and a combined SNR of 47 per pixel after 5 hours of observations (11 orbits of the binary). To reduce the ESI data, we used the MAuna Kea Echelle Extraction (MAKEE) pipeline, which does bias and sky subtraction, flat-fielding, removal of cosmic rays, spectral extraction, and wavelength calibration. 

\begin{figure}
\centering
\includegraphics[width=3.2in, clip=true, trim=0.3in 2.2in 0.6in 1.4in]{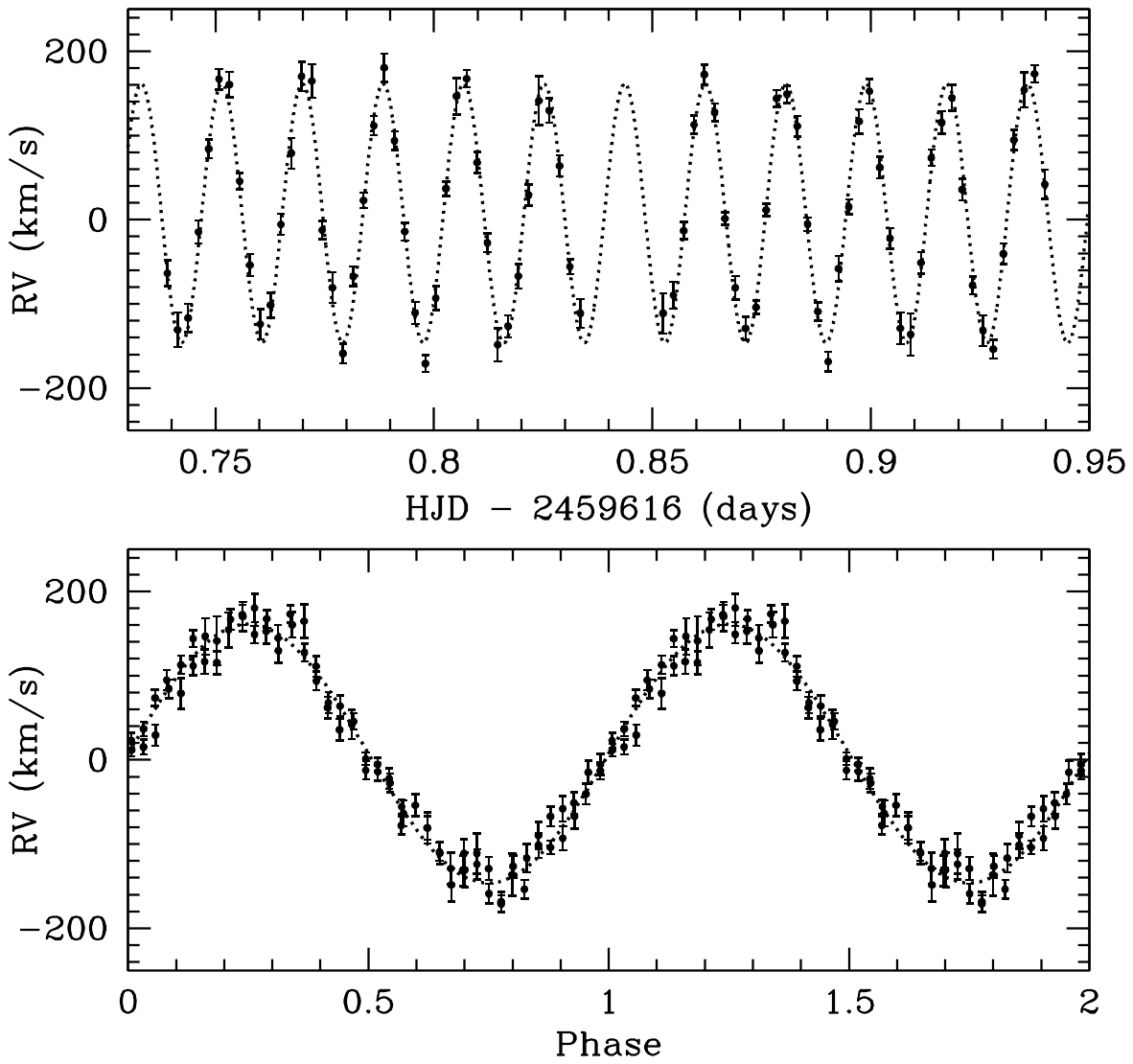}
\caption{Keck II/ESI radial velocities of J0634+3803 based on the combination of H$\alpha$, H$\beta$, and H$\gamma$ lines. The dotted line shows the best-fitting circular orbit. The bottom panel shows all of the data points phased with the best-fit period.}
\label{fig:5}
\end{figure}

This setup enables us to use all of the visible Balmer lines for velocity measurements. Figure~\ref{fig:5} shows the radial velocity measurements for J0634+3803 based on a combination of H$\alpha$, H$\beta$, and H$\gamma$ lines as black-filled circles; the black dotted line is the best-fitting circular orbit with $P= 1592.87 \pm 0.95$ s and $K= 153.6 \pm 2.1$ km s$^{-1}$.
These should be compared with $P=1591.4 \pm 28.9$ s and $K=132.1 \pm 6.0$ km s$^{-1}$ reported by \citet{kilic2021}. 

\begin{figure}
\centering
\includegraphics[width=3.2in, clip=true, trim=0.3in 2.2in 0.4in 1.4in]{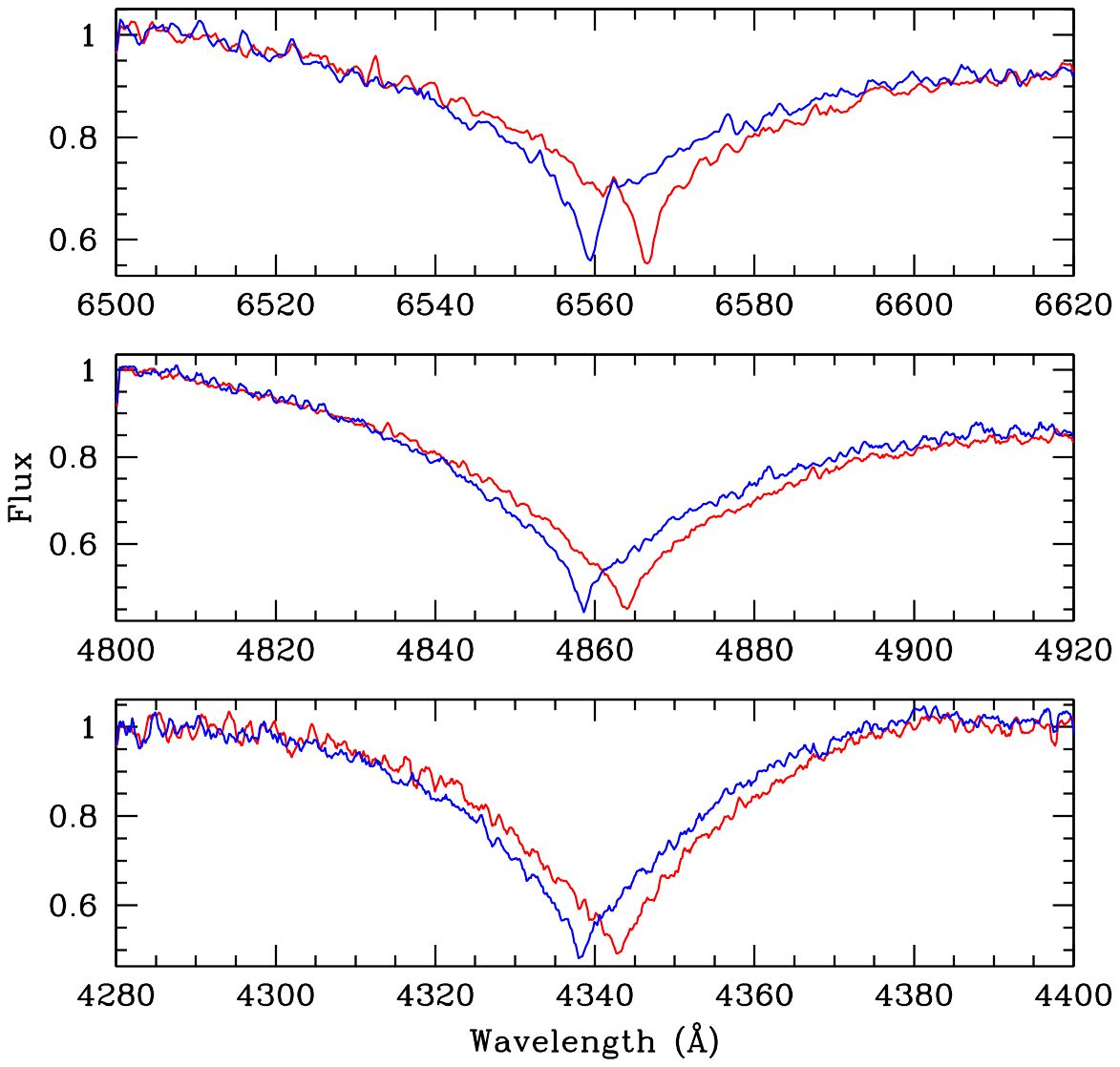}
\caption{Composite Keck II/ESI spectra for J0634+3803 during minimum (blue) and maximum (red line) radial velocities. The top, middle, and bottom panels show H$\alpha$, H$\beta$, and H$\gamma$, respectively.}
\label{fig:6}
\end{figure}

Keck II/ESI data significantly improve the period measurement and also reveal potential differences between the velocity semi-amplitudes measured at the MMT using the high-order Balmer lines (H$\gamma$ and higher) and at Keck using H$\gamma$ through H$\alpha$. In fact, even within the Keck II/ESI dataset, there are differences between the radial velocity semi-amplitudes measured from different lines. We obtain $K= 140.2 \pm 4.9$, $149.5 \pm 3.3$, and $163.6 \pm 3.5$ km s$^{-1}$ from H$\gamma$, H$\beta$, and H$\alpha$, respectively. The former measurement from H$\gamma$ ($K = 140.2 \pm 4.9$ km s$^{-1}$) is consistent with the velocity semi-amplitude measured from the MMT Blue Channel. This suggests that the contamination from the cooler secondary likely impacts our radial velocity measurements, shifting it slightly, depending on which H line is used. 

Based on a joint analysis of the photometry and spectroscopy data on J0634+3803, \citet{kilic2021} found that the cooler secondary in the system likely contributes significant flux in the optical region. However, there is a strong degeneracy between the physical parameters of the primary and secondary WDs (see their Figure 6). Basically, there are many possible combinations of $\log{g}$ and $T_{\rm eff}$ for the primary and secondary stars that can replicate the observed spectral energy distribution. Depending on the exact flux contribution and the temperature of the secondary star, it could impact the different Balmer lines by different amounts. They also found evidence of low-level ($0.34\pm0.12$\%) photometric variability at the orbital
period, but did not detect any evidence of ellipsoidal variations.

We searched for absorption features from the secondary star in our ESI data. Figure~\ref{fig:6} shows our composite Keck II/ESI spectra for minimum (blue) and maximum (red) radial velocities observed. The top, middle, and bottom panels show H$\alpha$, H$\beta$, and H$\gamma$, respectively. There is no clear evidence of a secondary line in these spectra. There may be a weak secondary H$\alpha$ feature in these composite spectra, but it is difficult to confirm, given the noise. Therefore, we cannot reliably constrain the flux contribution from the secondary star, and we adopt the velocity semi-amplitude of $K= 153.6 \pm 2.1$ km s$^{-1}$ from the combination of H$\alpha$, H$\beta$, and H$\gamma$ as presented in Figure \ref{fig:5}. 

Solving the binary mass function (see Equation~\ref{eq:1}) for our measurements of orbital period, velocity semi-amplitude, and mass constraints from \citet{kilic2021}, we revise the inclination measurement for J0634+3803 to $i$= 43$^{+7.0}_{-5.6}$ deg. Using the new information, we estimate the Doppler beaming amplitude to be $\sim 0.082$\%, which is below the detection limit in the light curve presented by \citet{kilic2021}. All of the updated parameters for this system are summarized in Table~\ref{tab:1}.

\begin{table}
    \centering
    \def\arraystretch{1.50}
    \setlength{\tabcolsep}{5.0pt}
    \caption{System Parameters}
    \begin{tabular}{l||cc}
             &  J2322+0509 & J0634+3803  \\
       \hline\hline
        R.A. & 23:22:30.20 & 06:34:49.92 \\
        Decl. & +5:09:42.06 & +38:03:52.2 \\
        $d$ (pc) & $865 \pm 168$ & $435^{+16}_{-15}$ \\
        $g$ (mag) & $18.578 \pm 0.008$ & $17.001 \pm 0.010$ \\
        $E(B-V)$ (mag) & 0.062 & 0.153 \\
        $P$ (s) & $1201.43 \pm 5.75$ & $1592.87 \pm 0.95$ \\
        $K$ (km s$^{-1}$) & $151.72 \pm 5.79$ & $153.6 \pm 2.1$ \\
        $\gamma$ (km s$^{-1}$) & $3.08 \pm 4.29$ & $7.40 \pm 1.40$ \\
        $T_{\text{eff,1}}$ (K) & $19870 \pm 300$ & $27300^{+4000}_{-2900}$ \\
        $\log g_1$ (cm s$^{-2}$) & $7.00 \pm 0.10$ & $7.46^{+0.28}_{-0.22}$ \\
        $M_1$ (M$_\odot$) & $0.31 \pm 0.02$ & $0.45^{+0.07}_{-0.06}$ \\
        $T_{\text{eff,2}}$ (K) & $9570 \pm 800$ & $10500^{+300}_{-200}$ \\
        $\log g_2$ (cm s$^{-2}$) & $7.15 \pm 0.20$ & $6.72^{+0.19}_{-0.13}$ \\
        $M_2$ (M$_\odot$) & $0.29 \pm 0.05$ & $0.21^{+0.03}_{-0.02}$ \\
        $i$ (°) & $25_{-3.0}^{+4.5} $ & $43_{-5.6}^{+7.0}$ \\
        \hline\hline
    \end{tabular}
    \label{tab:1}
    \vspace{10mm}\\
    \textbf{Note.} $T_{\rm eff}$, $\log{g}$, and masses for J0634+3803 are from \cite{kilic2021}.
\end{table}

\vskip 8mm

\section{Discussion}
\label{sec:4} 

\subsection{GW Parameters and Detection}
\label{sec:4.1.1}

To study the detectability of these systems with LISA, we employed the \texttt{LDASOFT}\footnote{\url{https://tlittenberg.github.io/ldasoft/html/index.html}} package \citep[][]{littenber2020} and the \texttt{vgb$\_$mcmc} sampler, which is a parallel-tempered Markov Chain Monte Carlo sampler to model the GW signals. We used $\delta$-function priors on the binary orbital period ($P$) and sky position ($\lambda, \beta$), fixing these parameters based on the EM observations. We sample the remaining parameters, such as GW amplitude and inclination, using uniform priors, where the initial value for inclination is set to the inclination derived from the EM measurements.

The sampler systematically marginalizes over nuisance parameters, including the first derivative of the GW frequency ($\dot{f}$), polarization angle ($\Psi$), and initial phase ($\phi_0$), ensuring that uncertainties in these do not bias the recovery of key system properties. The simulated data generated by \texttt{vgb$\_$mcmc} incorporates stationary Gaussian noise consistent with the LISA Data Challenge 2a\footnote{\url{https://lisa-ldc.lal.in2p3.fr/challenge2a}} \citep{LDCv2} noise models alongside an astrophysical foreground arising from unresolved Galactic binaries, following the framework of \cite{c&r2017}. This approach enables a reasonable assessment of LISA's capability to detect and characterize these sources.

\begin{figure*}
\centering
\includegraphics[width=3.5in]{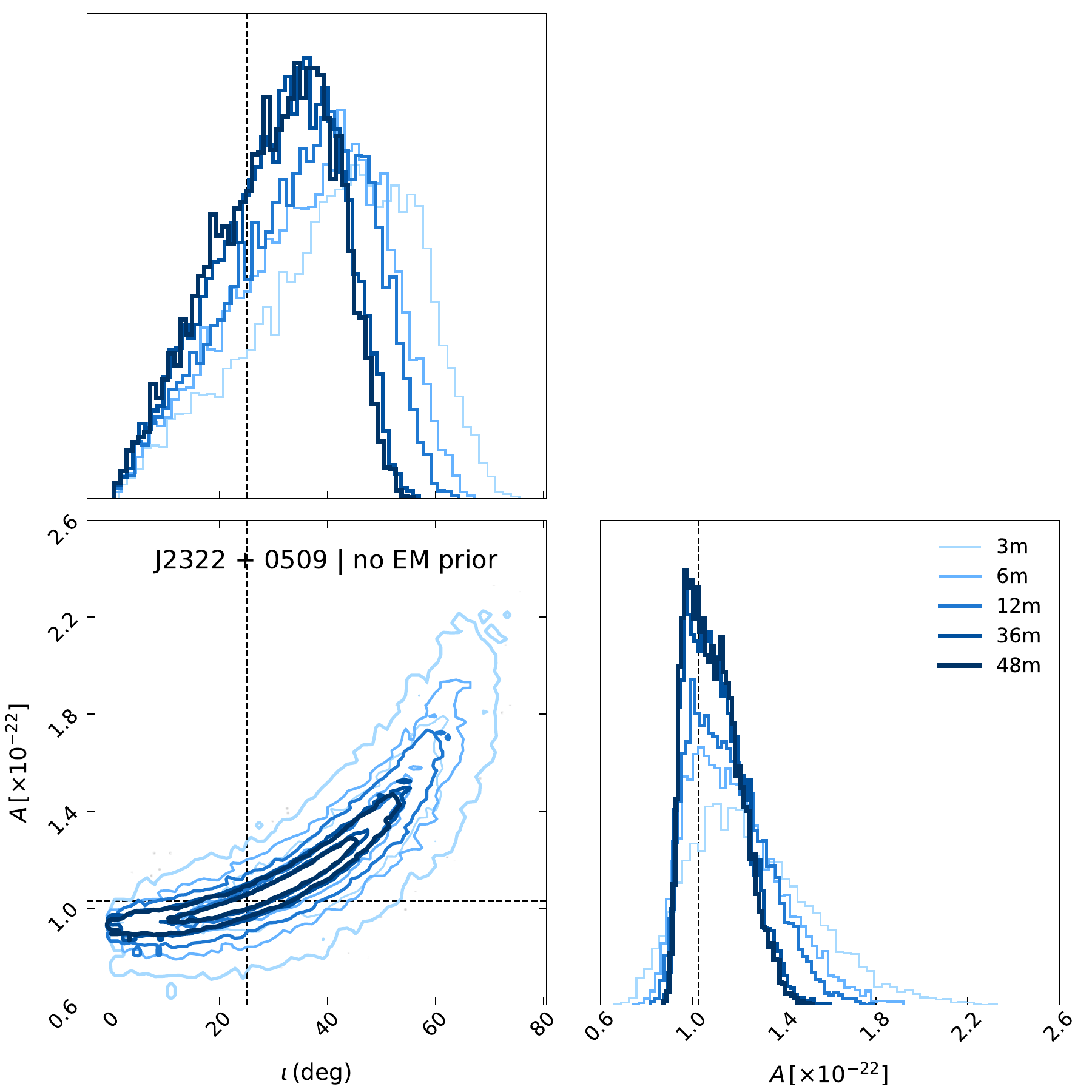}
\includegraphics[width=3.5in]{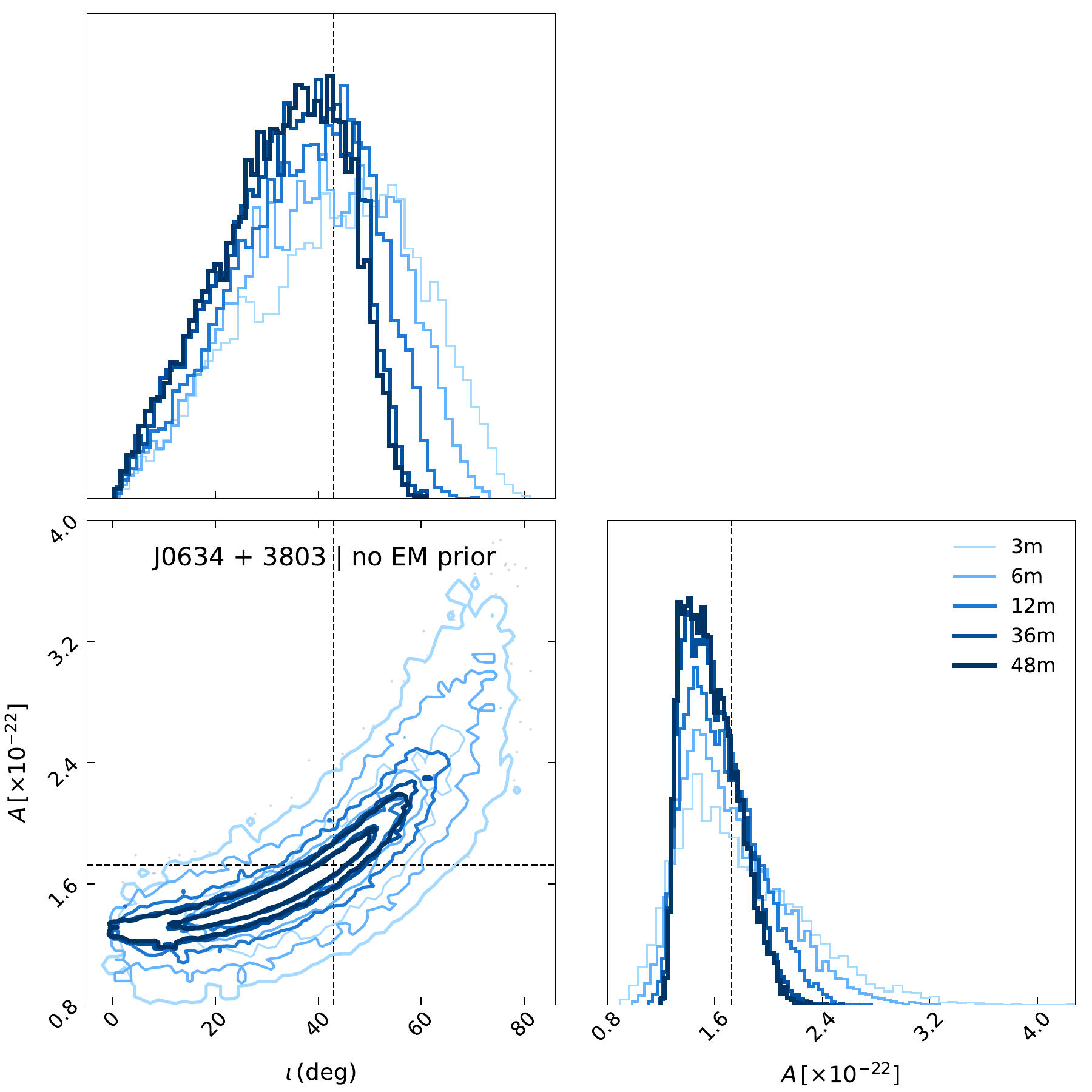}
\includegraphics[width=3.5in]{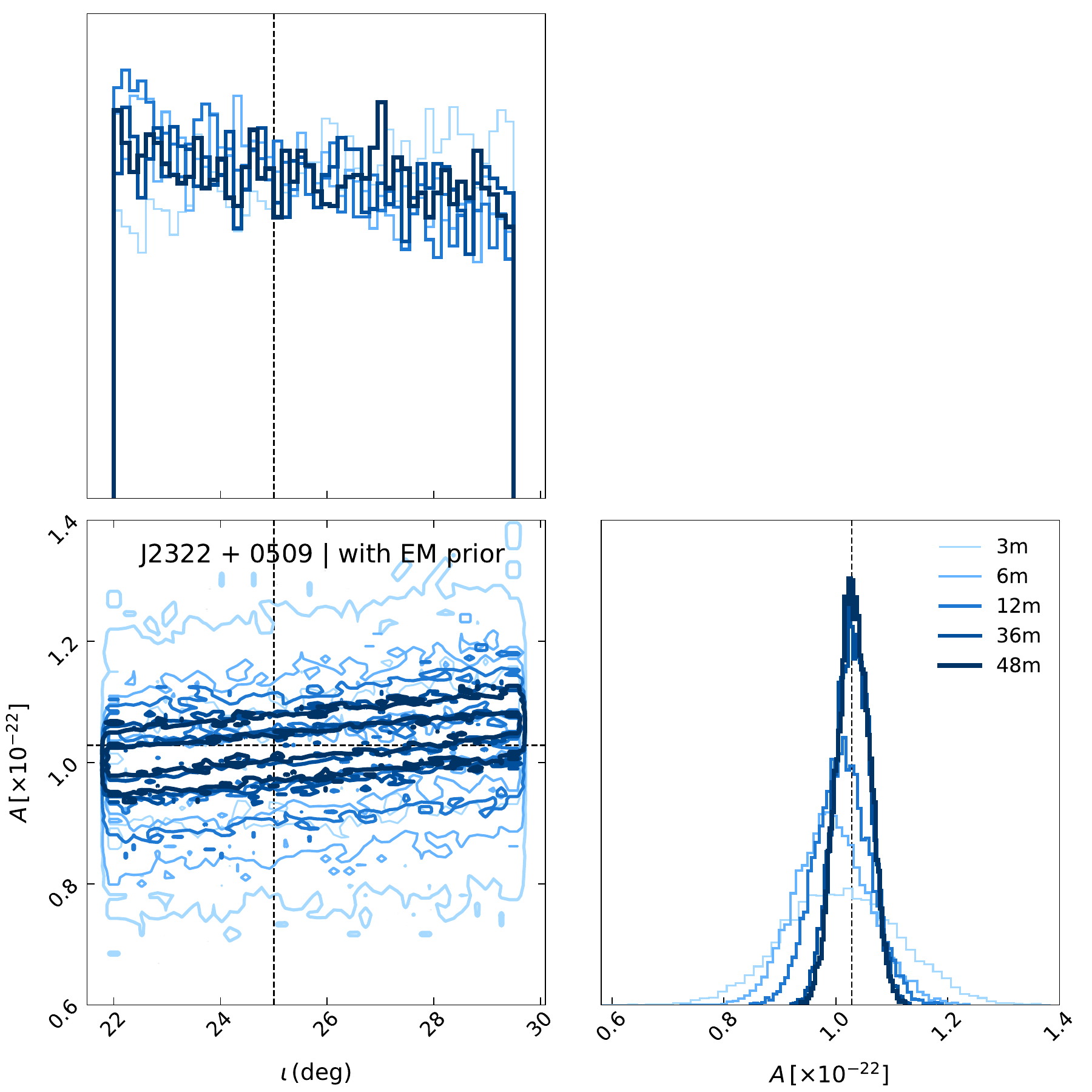}
\includegraphics[width=3.5in]{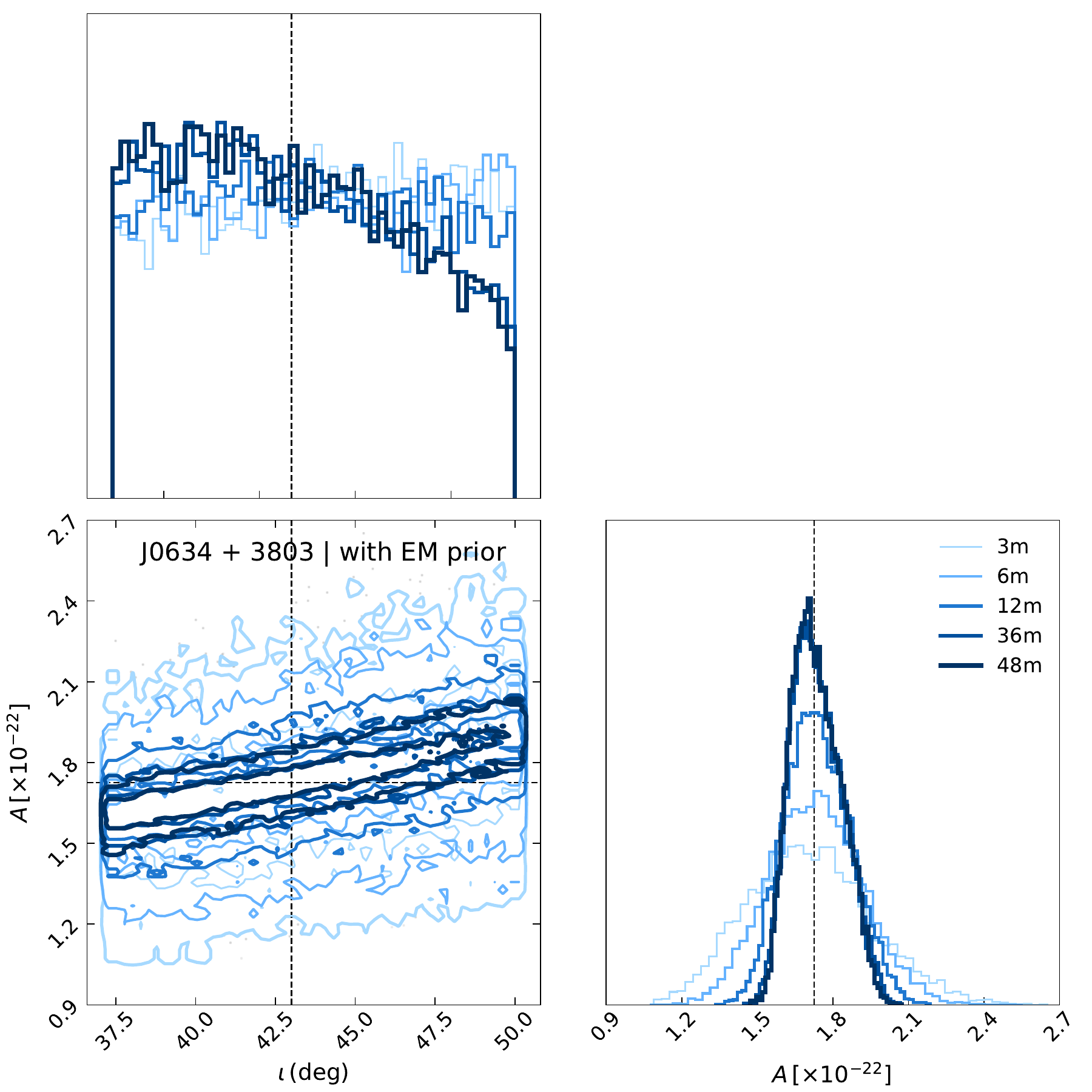}
\caption{Posteriors for binary inclination vs GW amplitude for J2322+0509 (left panels) and J0634+3803 (right panels) for 3, 6, 12, 36, and 48 months of operation time. The top panels show the targeted simulation without an EM prior on inclination, and the bottom panels show the posterior distribution when accounting for an EM inclination prior. The dashed lines indicate the injected inclination and GW amplitude in the simulations.}
\label{fig:7}
\end{figure*}

We simulated LISA science operation times of 3, 6, 12, 36, and 48 months. We explore the 2D posterior distribution in the GW amplitude-inclination plane to assess the detectability of each target by studying the different contour plots. A closed contour in this plane means that the GW amplitude is constrained away from the minimum value allowed, and therefore, the source will be detectable \citep[see Section 4 in][]{kupfer2024}. 

Figure~\ref{fig:7} shows the posterior inclination and GW amplitude distributions for J2322+0509 (left panels) and J0634+3803 (right panels) for 3, 6, 12, 36, and 48 months. Black dashed lines mark the injected values in the simulations for inclination and GW amplitude. The top and bottom panels show the 1$\sigma$ and 3$\sigma$ contours without and with the EM-priors on inclination, respectively. For the EM-prior case, \texttt{LDASOFT} assumes a uniform prior within the $\pm1\sigma$ uncertainties derived
from EM observations. 

For J2322+0509, the contours narrow down with increasing operation time for both (with and without EM-priors) cases, as expected. Figure~\ref{fig:7} top-left panel shows the results from the simulations when no EM prior is used: we observe closed contours after 6 months of observations where the detection is above $5\sigma$ significance. After 48 months of observations with LISA, the contours show significant improvement where the amplitude posterior distribution peaks and narrows around the injected $\mathcal{A}$ value with a fractional error $\sigma_\mathcal{A}/\mathcal{A} \approx 0.11$. However, the inclination posterior distribution peaks away from the injected value and is closer to $31\degree$ even after the full 4-year LISA mission lifetime, with an estimated precision of $\Delta i \approx 12\degree$. When comparing these results to \cite{kupfer2024}, who used an unconstrained inclination of $60\degree$ and \cite{brown2020} parameters ($\sigma_\mathcal{A}/\mathcal{A} \approx 0.30$ and $\Delta i \approx 36\degree$), we see a considerable improvement in both posterior distributions for our updated characterization of the secondary member of the system and the binary inclination angle.

Figure~\ref{fig:7} bottom-left panel shows the results from the same simulations, but with an EM-prior on inclination included. In this case, the parameter space is more tightly constrained, and the contours close faster. The source is already marginally detectable after a few months, where the amplitude posterior distribution spans a shorter amplitude space than the 48-month observation discussed above. Longer operation times refine the amplitude and inclination, achieving $\sigma_\mathcal{A}/\mathcal{A} \approx 0.03$ and $\Delta i \approx 3\degree$ by the end of the LISA mission, which is comparable to the EM uncertainties.

Similarly, Figure~\ref{fig:7} top-right panel shows the contours for the J0634+3803 system without the EM priors on inclination. The contours shrink considerably after 3 months of observations, reaching detectability conditions between 6 to 12 months, where the signal is above $5\sigma$ significance. However, both the amplitude and the inclination posterior distributions peak slightly to the left of the injected values, with $\mathcal{A}\approx1.54 \times 10^{22}$ and $i\approx34\degree$. At the end of the mission, the amplitude fractional error for this system is $\sigma_\mathcal{A}/\mathcal{A} \approx 0.14$, and the uncertainty in inclination is $\Delta i \approx 13\degree$, which slightly improves the parameters presented in \citet[][$\sigma_\mathcal{A}/\mathcal{A} \approx 0.19$ and $\Delta i \approx 28\degree$]{kupfer2024}. Adding the EM-prior on inclination leads to a significant improvement in the GW detection time-frame, as seen in the bottom-right panel of Figure~\ref{fig:7}. The contour is closed after 3 months, and the source is detectable with the amplitude and inclination posterior distribution peaking at the injected values. In this case, the amplitude fractional error improves to $\sigma_\mathcal{A}/\mathcal{A} \approx 0.06$ and $\Delta i \approx 4\degree$ after 48 months of observations.

\begin{figure}
\centering
\includegraphics[width=3.3in]{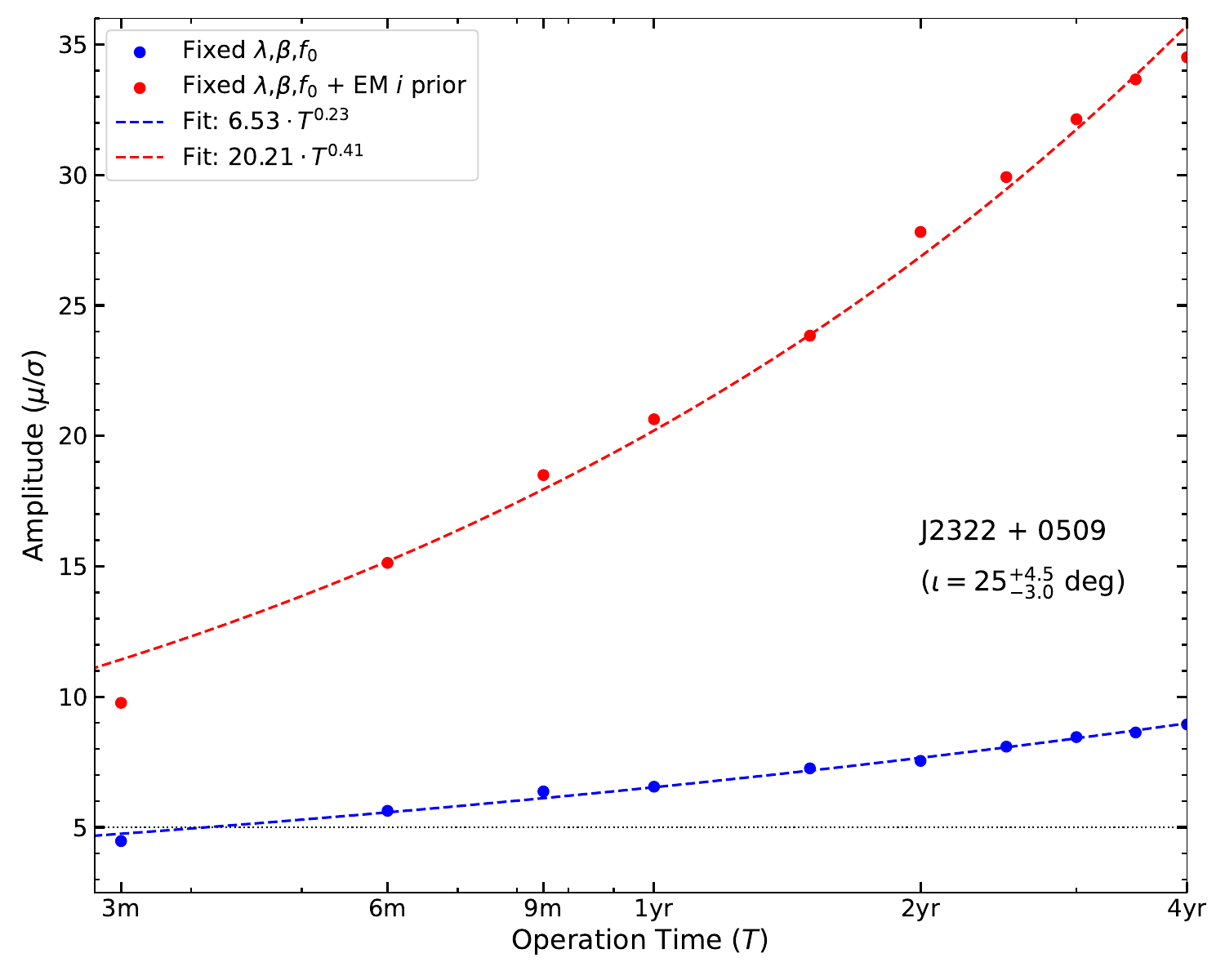}
\includegraphics[width=3.3in]{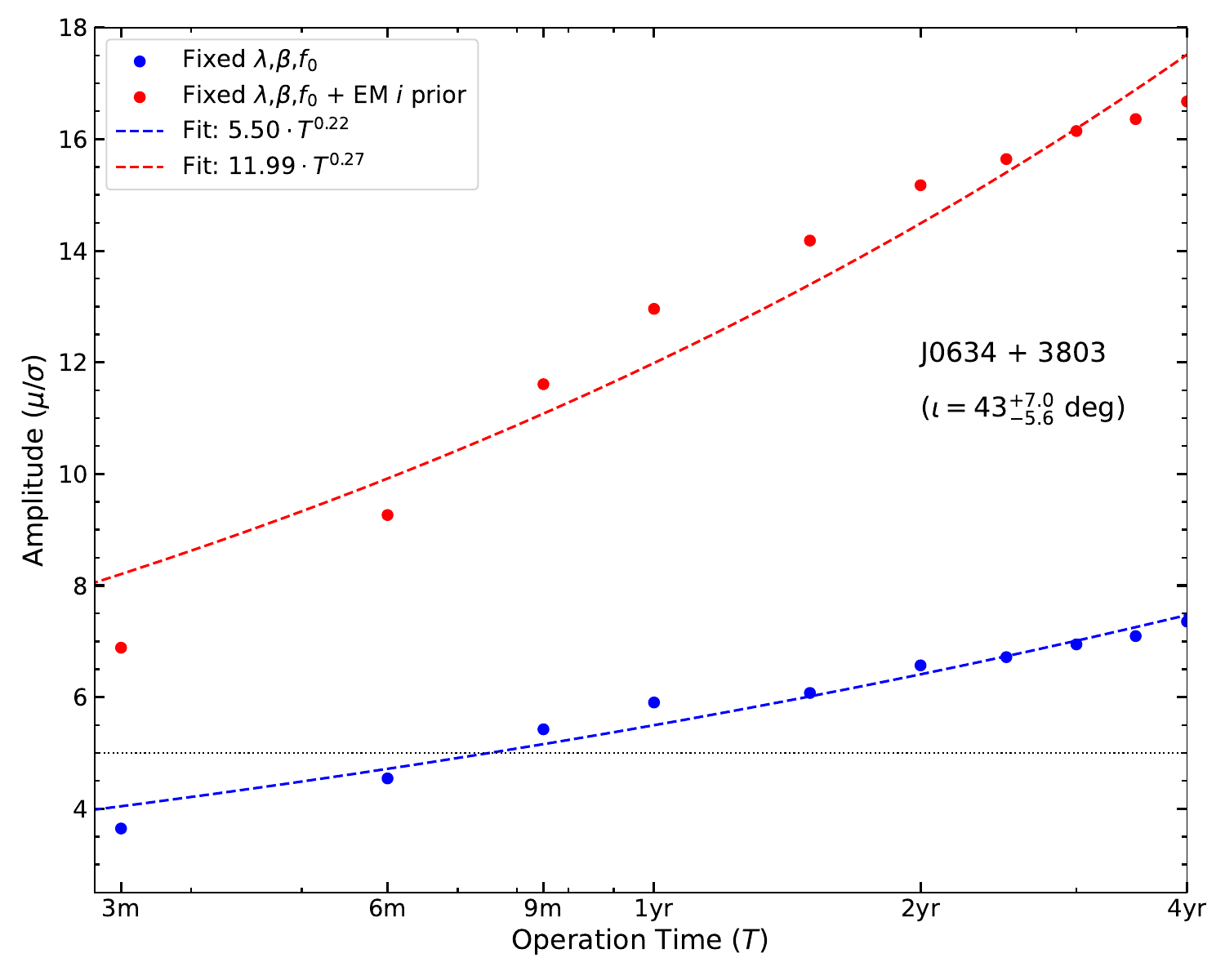}
\caption{Mean over standard deviation of the amplitude posteriors for J2322+0509 (top panel) and J0634+3803 (bottom panel) as a function of the mission duration ($T$). Red points are the values of $\mu/\sigma$ obtained by running GW simulations with fixed sky position ($\lambda, \beta$) and frequency ($f_0$) and adding an EM inclination prior; the blue points do not account for the latter. The red and blue dashed lines show the best fit to our data, and the dotted line shows our detection threshold at $\mu/\sigma$=5. In both cases, prior knowledge of the binary inclination through EM observations reduces the time of detection considerably.}
\label{fig:8}
\end{figure}

For operation times under 3 months, we obtain significantly different amplitude posterior distributions for J0634+3803. This is likely due to the mis-modeling of the data, which results in the contamination of the recovered signal by other, unknown galactic sources \citep[see Section~5 in][]{littenberg2024}, but including those additional noise sources is beyond the scope of this paper.

\subsection{Quantifying the Impact of EM-priors on Inclination}
\label{sec:4.2}

To quantify the impact of incorporating EM priors on inclination in GW analysis, we analyzed how the ratio of the mean amplitude to its standard deviation ($\mu/\sigma$) evolves with LISA's operation time. Figure~\ref{fig:8} presents $\mu/\sigma$ as a function of operation time for J2322+0509 (top panel) and J0634 $+$ 3803 (bottom panel). The red points mark the $\mu/\sigma$ values for the GW simulations for fixed sky position ($\lambda, \beta$), frequency ($f_0$), and an EM prior on inclination, and the blue points indicate the case without the latter. The red and blue dashed lines show the best fit power laws to the data, and the dotted line at $\mu/\sigma$ = 5 indicates a $5\sigma$ significant detection. In both cases, $\mu/\sigma$ increases with longer operation times as expected. More importantly, there is a substantial increase in this parameter when the EM priors on inclination are included in the GW analysis. For early detections ($\sim 3$ months), the extra information from EM observations increases $\mu/\sigma$ by a factor of two in both systems. In addition, after the full LISA mission, $\mu/\sigma$ increases by a factor of four for J2322+0509 and a factor of three for J0634+3803.

For J2322+0509, the red dashed line shows a higher initial amplitude and a faster growth rate ($T^{0.41}$) with time when the EM prior on inclination is included. Without the EM prior on the inclination, the growth rate is considerably slower ($T^{0.23}$). Hence, without the EM prior on inclination, a significant detection of this system's GW would take longer. In contrast, the longer period and higher inclination system J0634+3803 exhibits a less pronounced difference between the two cases ($T^{0.27}$ and $T^{0.22}$, respectively). Nevertheless, incorporating the inclination prior results in the source becoming detectable $\sim6$ months earlier than without the prior. It is important to note that for J0634+3803, the component masses are poorly constrained \citep[see][]{kilic2021}, which introduces uncertainties in the chirp mass estimation. This contributes to broader amplitude distributions in the GW simulations, affecting our detectability assessment.

\citet[][]{shah2012} demonstrated that EM-priors on inclination can improve the amplitude uncertainty up to a factor of $\sim6.5$ after 2 years of observations. Moreover, \citet[][]{shah2013} determined that along with the inclination, knowing the sky position of a GW source from EM observations can reduce the uncertainty of the GW parameters up to a factor of $\sim2$. Including EM-priors on both sky position and inclination can improve the uncertainty in amplitude by a factor $\geq40$. 

Since we are interested in studying GW detectability of known short-period DWD systems, all of our simulations use $\delta$-function priors on the binary orbital period and sky position as measured from EM observations. Hence, we cannot make an exact comparison with the \citet{shah2013} results. However, \citet{finch2023} took a similar approach as ours, where they found that the EM-prior
on inclination leads to an improvement of a factor of $\sim2.4$ in GW amplitude measurement after 2 years of observation time for a face-on system ($i=15\degree$).  In our simulations, an EM-prior on inclination improves the GW amplitude measurement by a factor of $\sim3.7$ and $\sim2.3$ after 2 years of observation time for J2322+0509 and J0634+3803, respectively. Hence, precise EM constraints on binary inclination can significantly aid the GW amplitude measurements.

\section{Summary and Conclusions}
\label{sec:5}

J2322+0509 and J0634+3803 are two DWD binaries located at 865 and 435 pc, respectively. For J2322+0509, we were able to constrain the mass of both objects, with masses $\approx0.30~M_{\odot}$, thanks to the new HST/STIS UV spectra and the improved measurements of the period, radial velocity, and orbital inclination of the system. For J0634+3803, we improved the period, radial velocity, and orbital inclination measurements using new spectroscopic data from KECK II/ESI.

Our GW analysis using LDASOFT shows that both sources will be detectable in the early stages of the LISA mission when accounting for the EM inclination prior, which significantly improves the accuracy of the GW amplitude by a factor of 2–4 and shortens the detection time by several months. For J2322+0509, using an EM prior improves the fractional error in GW amplitude to $\approx3\%$ after 48 months of LISA observations. Similarly, for J0634+3803, the GW amplitude fractional error shrinks to $\approx6\%$. 

A limitation of our analysis is the assumption that the astrophysical foreground behaves as a stationary noise source and that additional uncertainties from overlapping signals are negligible. These factors could impact the detection and parameter estimation of GW binaries in the mHz frequency range. Future studies should address these challenges by incorporating a global fit approach for DWD systems to achieve a more realistic characterization of LISA's data and instrument response. Another caveat is that \texttt{LDASOFT} uses a uniform prior on inclination, where we adopt a limited range in inclination based on the $\pm1\sigma$ uncertainties from the EM measurements. Future work on the LISA detectability of short-period binary WDs would benefit from including a more realistic (e.g., Gaussian) prior on inclination. 

Given the lack of precise inclination measurements in non-eclipsing binaries, our results demonstrate that UV characterization of the hot primary WD can help constrain the system parameters and, therefore, provide a more accurate EM inclination measurement. Future ultraviolet facilities such as the Ultraviolet Explorer \citep[UVEX,][]{kulkarni21} will significantly enhance the ability to conduct multi-messenger studies of compact binaries. UVEX will offer wide-field UV imaging and rapid spectroscopic follow-up capabilities. While HST provides superior spectral resolution, UVEX’s design prioritizes sky coverage and prompt access, making it highly complementary for characterizing gravitational-wave sources as shown in this work. In addition, observations with JWST could enable direct spectroscopic or photometric analysis of the cooler companion, especially in the near-infrared where its flux contribution is more prominent. This could offer a unique opportunity to constrain the physical parameters of the secondary star, complementing UV and optical constraints.

Finally, the upcoming Gaia DR4 will provide improved parallax measurements, while new large-scale photometric surveys, i.e., Vera Rubin Observatory's Legacy Survey of Space and Time will expand the number of EM-detected GW binaries. Follow-up observations of these sources will be essential for maximizing LISA's scientific return and improving our understanding of binary evolution.

\section*{Acknowledgements}

We thank the referee for useful comments that helped us improved this manuscript. This work is supported in part by the National Aeronautics and Space Administration under grants 80NSSC24K0436, 80NSSC22K0479, and 80NSSC24K0380, the National Science Foundation under grant AST-2205736, the Smithsonian Institution, and the Natural Sciences and
Engineering Research Council of Canada (NSERC). A.B. is an NSERC Postdoctoral Fellow.

Based on observations with the NASA/ESA Hubble Space Telescope obtained from the Mikulski Archive for Space Telescopes (MAST) at the Space Telescope Science Institute, which is operated by the Association of Universities for Research in Astronomy, Incorporated, under NASA contract NAS5-26555. Support for Program number (HST-GO-16286.001-A) was provided through a grant from the STScI under NASA contract NAS5-26555.

This work was supported by a NASA Keck PI Data Award, administered by the NASA Exoplanet Science Institute. Data presented herein were obtained at the W. M. Keck Observatory from telescope time allocated to the National Aeronautics and Space Administration through the agency's scientific partnership with the California Institute of Technology and the University of California. The Observatory was made possible by the generous financial support of the W. M. Keck Foundation.

This work was co-funded by the European Union (ERC, CompactBINARIES, 101078773). Views and opinions expressed are, however, those of the author(s) only and do not necessarily reflect those of the European Union or the European Research Council. Neither the European Union nor the granting authority can be held responsible for them.

The authors wish to recognize and acknowledge the very significant cultural role and reverence that the summit of Maunakea has always had within the Native Hawaiian community. We are most fortunate to have the opportunity to conduct observations from this mountain.

\section*{Data availability}

The data underlying this article are available in the Mikulski Archive for Space Telescopes (MAST) and the Keck Observatory Archive. All the {\it HST}, data used in this paper can be found in MAST: \dataset[10.17909/35rz-vb77]{http://dx.doi.org/10.17909/35rz-vb77}.
The reduced spectroscopy data that support the findings of this study are available from the corresponding author upon reasonable request.




\bibliography{LISA_sources}{}
\bibliographystyle{aasjournal}

\label{lastpage}
\end{document}